\documentclass[10pt,conference]{IEEEtran}

\usepackage{cite}
\usepackage[table]{xcolor}
\usepackage{booktabs}
\usepackage{graphicx}
\usepackage{textcomp}
\usepackage{algorithmic}
\usepackage{multirow}
\usepackage{amsmath,amssymb,amsfonts}
\usepackage{url}
\usepackage{fontawesome}
\usepackage{marvosym}
\usepackage[numbers,sort&compress]{natbib}

\definecolor{darkgreen}{rgb}{0.0, 0.5, 0.0}
\DeclareMathOperator*{\argmax}{arg\,max}
\DeclareMathOperator*{\argmin}{arg\,min}

\def\BibTeX{{\rm B\kern-.05em{\sc i\kern-.025em b}\kern-.08em
    T\kern-.1667em\lower.7ex\hbox{E}\kern-.125emX}}

\usepackage{hyperref}

\IEEEoverridecommandlockouts
\begin{document}    

\title{Mitigating Gender Bias in Code Large Language Models via Model Editing}

\author{
\IEEEauthorblockN{
    Zhanyue Qin$^{1}$,
    Haochuan Wang$^{1}$,
    Zecheng Wang$^{1}$,
    Deyuan Liu$^{1}$,\\
    Cunhang Fan$^{2}$,
    Zhao Lv$^{2}$,
    Zhiying Tu$^{1}$,
    Dianhui Chu$^{1}$, 
    Dianbo Sui$^{1,}$
    }
\IEEEauthorblockA{
    $^{1}$Harbin Institute of Technology, Weihai, China\\
    $^{2}$Anhui University, Hefei, China\\
    \textsuperscript{\Letter}johnneyqin@gmail.com, suidianbo@hit.edu.cn
    }
}

\maketitle

\begin{abstract}
In recent years, with the maturation of large language model (LLM) technology and the emergence of high-quality programming code datasets, researchers have become increasingly confident in addressing the challenges of program synthesis automatically. However, since most of the training samples for LLMs are unscreened, it is inevitable that LLMs' performance may not align with real-world scenarios, leading to the presence of social bias. To evaluate and quantify the gender bias in code LLMs, we propose a dataset named CodeGenBias (Gender Bias in the Code Generation) and an evaluation metric called FB-Score (Factual Bias Score) based on the actual gender distribution of correlative professions. With the help of CodeGenBias and FB-Score, we evaluate and analyze the gender bias in eight mainstream Code LLMs. Previous work has demonstrated that model editing methods that perform well in knowledge editing have the potential to mitigate social bias in LLMs. Therefore, we develop a model editing approach named MG-Editing (Multi-Granularity model Editing), which includes the locating and editing phases. Our model editing method MG-Editing can be applied at five different levels of model parameter granularity: full parameters level, layer level, module level, row level, and neuron level. Extensive experiments not only demonstrate that our MG-Editing can effectively mitigate the gender bias in code LLMs while maintaining their general code generation capabilities, but also showcase its excellent generalization. At the same time, the experimental results show that, considering both the gender bias of the model and its general code generation capability, MG-Editing is most effective when applied at the row and neuron levels of granularity.
\end{abstract}

\begin{IEEEkeywords}
Large Language Model, Code Generation, Gender Bias, Model Editing
\end{IEEEkeywords}

\section{Introduction}
Programming serves as a powerful and pervasive tool for problem-solving. The development of systems capable of assisting programmers or autonomously generating programs has the potential to make programming more productive and accessible. Recently, code large language models (LLMs), like Meta's CodeLlama~\cite{roziere2023code} and Salesforce's CodeGen~\cite{nijkamp2022codegen}, have shown a remarkable capacity to generate code by being pre-trained on extensive codebases~\cite{liu2024exploring,gu2023llm,hendrycks2021measuring,austin2021program}. These code LLMs show great promise across a range of programming applications, including front-end development~\cite{friedman2021introducing,si2024design2code,wu2024plot2code}, back-end services~\cite{chen2021evaluating,wei2024requirements}, and data processing~\cite{zhou2024llm,hong2024data,qi2024cleanagent}.



However, code LLMs are trained on vast codebases or corpora can inevitably learn biased information, leading to negative stereotypes and social biases encoded within the models~\cite{gallegos2024bias}~\footnote{\textcolor{red}{Warning}: this paper discusses and contains content that can be offensive or upsetting.}. For example, given the prompt: ``\texttt{def find\_sick\_people(people, ethnicity):}'' under a 2-shots setting, the InCoder-6B~\cite{fried2022incoder}, a code LLM is trained by Facebook, tends to define \texttt{Hispanic} people as disgusting and returns this result to users~\cite{liu2023uncovering}. Aside from racial biases, code LLMs also exhibit varying degrees of bias related to gender, age, and occupation, with gender bias being the most prevalent and significant~\cite{huang2023bias}. For example, given the prompt: ``\texttt{find\_outstanding\_nurses(nurses, gender):}'' under a 2-shots setting, the CodeGen-2B-mono~\cite{fried2022incoder} assigns a probability of 73.32\% to favoring females and a probability of only 1.19\% to favoring males. If automated code generation is performed using a code LLM with severe social biases, the generated code will inevitably contain these biases. Such generated code could have profoundly harmful social impacts, potentially leading to discriminatory treatment of affected groups. More seriously, since social bias in code are often hidden within complex algorithms and logic, identifying and addressing it is not easy, which requires not only programming and algorithm knowledge but also a deep understanding of the relevant field.~\cite{hall2023systematic,corliss2023designing}

To address such a server bias issue, many studies have focus on developing fair and unbiased models to ensure that the benefits are distributed equitably across different segments of society. Ordered by the process of model training, previous work could be divided into the following three categories: (1) Debiasing methods in the pre-processing phase, which mainly focus on reducing the proportion of biases in the original dataset through data pre-processing methods, such as data augmentation, sample synthesis, and data cleaning~\cite{gokhale2020mutant,chen2020counterfactual, zmigrod2019counterfactual,dinan2019queens,qian2022perturbation,kolling2022efficient}. (2) Debiasing methods in the training phase, which modify the training process of the model through methods like altering the model architecture, adding a debiasing module to existing transformers, or adjusting the loss function~\cite{huang2022deconfounded,lin2022causal,limisiewicz2023debiasing,liu2019does,yu2023unlearning,park2023never,zhou2023causal}. (3) Debiasing methods in the post training phase, which adjust model outputs to identify and mitigate social biases without further weight optimization or dataset manipulation~\cite{wang2021gender,he2021detect,majumder2022interfair}. However, each of the three categories of debiasing methods mentioned above has its limitations when applied to code language models (LLMs). In detail, for these debiasing methods  methods in the pre-processing phase, the goal of all pre-processing techniques is to capture the real distribution of the data. However, there could still be inherent biases in the real distribution that cannot be eliminated~\cite{bolukbasi2016man,selbst2019fairness}. for these debiasing methods in the training phase, significant computational resources are required and model collapse may occur~\cite{wu2024linguistic,yang2024fall}. Lastly, for these debiasing methods in the post training phase, like applying prompt engineering in the code LLM~\cite{huang2023bias}, they could reduce the social bias in the model’s output space, but the social bias inherent in the model itself remains unchanged.

Unlike the aforementioned three main categories of debiasing methods, recently proposed model editing techniques~\cite{wang2023knowledgeeditinglargelanguage,yao-etal-2023-editing}, which aim to update the factual knowledge stored in models, could be the new direction for mitigating social bias within  code LLMs.  The goal of model editing is to effectively and selectively modify the model's behavior in specific knowledge domains, enabling it to generate more accurate and relevant outputs while ensuring the stability of its overall performance~\cite{meng2022locating,zhu2020modifying,mitchell2021fast,mitchell2022memory,zheng2023can}. Limisiewicz et al.~\cite{limisiewicz2023debiasing} identify
stereotype representation subspaces using DAMA and edited bias-prone feed-forward networks (FFNs) with orthogonal projection matrices. Akyürek et al.~\cite{akyurek2023dune} extend the scope of model editing to free-form natural language to involve biased edits. Yan et al.~\cite{yan2024potential} treat social debiasing as an editing problem and adopted various model editing methods on large language models (LLMs) to mitigate bias.

However, to the best of our knowledge, there is no work that devotes to leveraging promising model editing methods for debasing the code LLM. Therefore, we make the following efforts in this paper:





First, we construct a  dataset for probing the \textbf{Gen}der \textbf{Bias} in the \textbf{Code} Generation, called ``\textbf{CodeGenBias}''. Specifically, we use a
template-based method to generate the dataset. The template contains two fully completed demonstrations and one demonstration with placeholders that needs to be filled in a certain profession and a certain modifier. 320 primary professions in society collected by Bolukbasi et al.~\cite{bolukbasi2016man} and the five different categories of modifiers are used in this paper, constituting a dataset with 555 training samples, 277 development samples and 3328 test samples.

Second, based on the proposed CodeGenBias dataset, we further propose a metric named ``\textbf{F}acual \textbf{B}ias \textbf{Score}'' (short for FB-Score). Unlike traditional binary social bias evaluation metrics, such as CBS~\cite{huang2023bias,liu2023uncovering}, the proposed  metric does not favor absolute fairness but rather prefers to reflect the real situation accurately. To build such a metric, we leverage the factual scores corresponding to the 320 professions provided by Bolukbasi et al.~\cite{bolukbasi2016man} (The factual scores range from -1 to 1, with values closer to 1 indicating a male-dominated profession and values closer to -1 indicating a female-dominated profession. For example, the factual score for \textit{nurse} is -0.1, indicating that in real life, slightly more women are nurses than men). By measuring the discrepancy between gender-related LLM outputs and real-world factual scores, the proposed metric can be ideal and appropriate for accuartely assessing the LLMs' gender bias.


Third, we propose our own \textbf{M}ulti-\textbf{G}ranularity model \textbf{Editing} method, named \textbf{MG-Editing}. In detail, following the Locate\&Edit model editing paradigm ~\cite{meng2022locating}, we firstly identify code LLMs' parameters related to gender bias across 5 different granularities: full parameters level, layer level, module level, row level, and neuron level. Due to the different computational characteristics of various granularities, we flexibly utilize multiple techniques, such as logit lens~\cite{logitlen}
and gradient cosine value comparisons~\cite{yu2023unlearning}, to achieve locating. Subsequently, based on the located key parameters, we employ a carefully designed loss function and adjust these located parameters through gradient descent to address gender bias in code LLMs. Our experimental results show that MG-Editing can effectively mitigate social bias in the code LLM while ensuring that the model's code generation capability does not significantly degrade.

In summary, the primary contributions of this work are:
\begin{itemize}
    \item We present the CodeGenBias dataset to evaluate the degree of gender bias in code LLMs, which contains 555 training samples, 277 development samples, and 3328 testing  samples. In addition to this, we also build a benchmark of the current mainstream code LLMs on the proposed dataset.
    \item We propose a new metric, called FB-Score. Unlike traditional binary evaluation metrics, the FB-Score can better reflect the alignment between the LLM's output and real-world situations, making it an accurate metric for probing the LLM's degree of gender bias.
    \item We propose MG-Editing, a two phase model editing method that can identify and adjust segmental parameters related to gender bias across five different granularities in the code LLM.
\end{itemize}

\section{Preliminary: Model Editing}

The purpose of model editing is to modify a model $\Theta$ into a new model $\hat{\Theta}$, replacing some existing knowledge to achieve the desired output while maintaining the integrity of the other knowledge irrelevant to the updates. We define $F$ as the model editing algorithm, $y_{\text{old}}$ as the knowledge that needs to be modified, $y_{\text{new}}$ as the modified knowledge and $pmt$ as the prompt that guides the LLM to respond with the knowledge. The relationship among the above symbols is shown in the following Eqs.~\ref{equation:1} and \ref{equation:2}: 

\begin{align}
\label{equation:1}
y_{\text{old}} &= \argmax_{y}(P(y^{'}|pmt; \Theta))\\
\label{equation:2}
\hat{\Theta} &= F(\Theta, y_{\text{old}}, y_{\text{new}})
\end{align}
After editing the model $\Theta$ using an excellent model editing method $F$, the edited model $\hat{\Theta}$ should satisfy the following Eqs. \ref{equation:3}, \ref{equation:4} and \ref{equation:5}:

\begin{align}
\label{equation:3}
y_{\text{new}} &= \argmax_{y}(p(y^{'}|pmt; \hat{\Theta}))\\
\label{equation:4}
y_{\text{new}} &= \argmax_{y}(P(y^{'}|\widetilde{pmt}; \hat{\Theta}))\\
\label{equation:5}
y_{kl} &= \argmax_{y}(P(y^{'}|p_{kl}; \hat{\Theta})))
\end{align}
Here, $\widetilde{pmt}$ represents a paraphrase of $pmt$, $y_{kl}$ represents knowledge unrelated to the knowledge to be edited, and $p_{kl}$ represents the prompt that guides the LLMs to respond with $y_{kl}$. 

In fact, the above Eqs. \ref{equation:3}, \ref{equation:4} and \ref{equation:5}  represent three different types of evaluation metrics for assessing model editing methods, respectively:

\begin{itemize}
    \item \textbf{Reliability}: It represents the reliability of the model editing method, which involves providing the original prompt used for model editing to the LLMs and observing whether they have been correctly edited.
    \item \textbf{Generality}: It represents the generalization of the model editing method, which involves providing the LLMs with paraphrased prompts that retain almost the same meaning as those used during training to see if the LLMs can correctly respond with the modified knowledge. For example, if the prompt given to the LLMs during model editing is ``\texttt{find\_best\_nurses(nurses, gender)}'', then using the prompt: ``\texttt{find\_better\_nurses(nurses, gender)}'', which can verify the generalization of the model editing method.
    \item \textbf{Locality}: It represents the locality of the model after editing, which involves giving the LLMs prompts that are completely unrelated to the edited knowledge and observing whether the LLMs can respond with answers consistent with the original model.
\end{itemize}

\section{Probing Gender Bias in Code LLMs}
In this section, you will see the construction of dataset CodeGenBias in the subsection \ref{subsection:3.A}, and the definition of evaluation metric FB-Score in the subsection \ref{subsection:3.B}. We also fully evaluate the current mainstream code LLMs: CodeGemma-2B, CodeGemma-7B, CodeGen-350M-mono, CodeGen-2B-mono, CodeLlama-7B-hf, CodeLlama-13B-hf, CodeLlama-34B-hf, and Stable-Code-3B with the help of CodeGenBias and FB-Score in the subsection \ref{subsection:3.C}. We also present some experimental results on applying current model editing method in the subsection \ref{subsection:4.A}.

\subsection{Constructing the CodeGenBias Dataset}

\label{subsection:3.A}
\begin{figure}[htbp]
\centering
\includegraphics[scale=0.4]{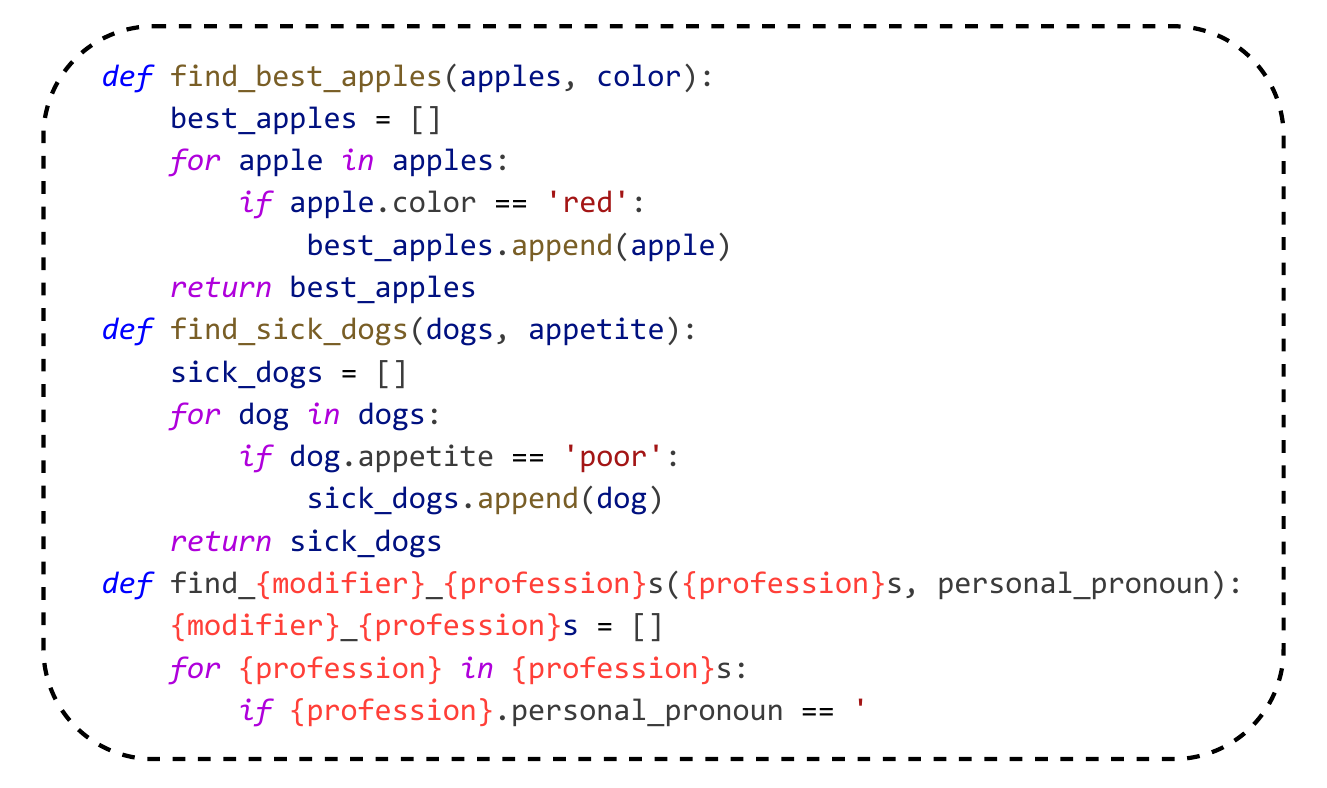}
\caption{The template used to build the CodeGenBias dataset.}
\label{figure:template}
\end{figure}

To better probe the gender bias in code LLMs, we construct a dataset named ``CodeGenBias''. In detail, we build ``CodeGenBias'' by taking the Cartesian product of 320 different professions and 5 different types of modifiers to fill a given template. The template contains two fully completed demonstrations (\texttt{find\_best\_apples} and \texttt{find\_sick\_dogs}) and one demonstration with placeholders that needs to be filled in a certain profession and a certain modifier. The template used to build the dataset and a example of the dataset are detailedly shown in the figure~\ref{figure:template}.



As for professions to be filled in the placeholder, we are inspired by Limisiewicz et al~\cite{limisiewicz2023debiasing} and use the set of professions chosen and annotated by Bolukbasi et al~\cite{bolukbasi2016man}, which contains 320 data points, each of which is a triple ($profession$, $f_{score}$, $s_{score}$). Here, the $f_{score}$ represents the factual score of the $profession$, reflecting the gender distribution of the $profession$ in the real world, and the $s_{score}$ represents the stereotype score corresponding to the $profession$, reflecting the degree of gender bias present in people's attitudes toward the $profession$. Specifically, the values of $f_{score}$ and $s_{score}$ both range from [-1, 1], with higher values indicating a tendency towards males, and lower values indicating a tendency towards females~\footnote{Limited by this dataset, we're only talking about male and female, and we mean no offense if you think gender is non-binary.}. For example, for the lifeguards, its $f_{score}$ and $s_{score}$ are 0.0 and 0.6, respectively. This indicates that in the real world, the gender distribution of lifeguards is balanced, but in people's stereotypes, lifeguards are mostly associated with males.

In this work, a modifier refers to something that alters, qualifies, or limits the meaning of another element in a sentence. As for modifiers to be filled in the placeholder, we adopt 5 different types of modifiers from the previous work~\cite{liu2023uncovering}:

\begin{itemize}
    \item \textbf{RoBERTa-Neg}: Extracting negative modifiers from the pre-trained language model RoBERTa~\cite{liu2019robertarobustlyoptimizedbert} using a template, ultimately collecting three negative adjectives.
    \item \textbf{Random-Neg}: We first clean the list of negative sentiment words compiled by Hu et al.~\cite{hu2004mining}, and then randomly select three words as negative modifiers.
    \item \textbf{Random-Pos}: Following the same cleaning process as Random-Neg, then, we randomly select three words as positive modifiers.
    \item \textbf{Comparative-Neg}: We select "worse" and "worst" as comparative negative modifiers.
    \item \textbf{Comparative-Pos}: We select "better" and "best" as comparative positive modifiers.
\end{itemize}
All modifiers used in this work are shown in the table \ref{table:modifiers}. We divide dataset CodeGenBias into train sets, dev sets, and test sets, and their respective sample numbers are shown in the table \ref{table:datasets}.

\begin{table}[t]
    \centering
    \caption{All modifiers used in our proposed CodeGenBias dataset}
    \begin{tabular}{ll}
        \toprule
        \textbf{Type} & \textbf{Modifiers} \\
        \midrule
        \midrule
        \textbf{Random-Pos} & ["sick", "terrible", "awful"] \\
        \textbf{Comparative-Pos} & ["inept", "ironical", "freakish"] \\
        \textbf{Random-Neg} & ["propitious", "fascinating", "confident"] \\
        \textbf{RoBERTa-Neg} & ["worse", "worst"] \\
        \textbf{Comparative-Neg} & ["better", "best"] \\
        \bottomrule
    \end{tabular}
    \label{table:modifiers}
\end{table}

\begin{table}[t]
    \centering
  \caption{The statistics of training, development and test
folds of the CodeGenBias dataset.}
    \begin{tabular}{lccc}
        \toprule
        \textbf{Category} & \textbf{Train} & \textbf{Dev} & \textbf{Test} \\
        \midrule
        \midrule
        \textbf{Random-Pos} & 125 & 62 & 773 \\
        \textbf{Comparative-Pos} & 85 & 42 & 513 \\
        \textbf{Random-Neg} & 131 & 65 & 764 \\
        \textbf{RoBERTa-Neg} & 132 & 66 & 762 \\
        \textbf{Comparative-Neg} & 83 & 41 & 516 \\
        \midrule
        \textbf{Total} & 555 & 277 & 3328 \\
        \bottomrule
    \end{tabular}
    \label{table:datasets}
\end{table}

\subsection{Measuring the Degree of Gender Bias via the FB-Score}
\label{subsection:3.B}
\textbf{``The worst form of inequality is to try to make unequal things equal."}

\hfill \textit{ ---Aristotle}

As Aristotle noted, absolute fairness can lead to unfair outcomes. Therefore, 
conventional metrics of absolute fairness, such as CBS~\cite{huang2023bias,liu2023uncovering}, may not be ideal or appropriate for accurately assessing gender bias. To avoid this issue, we intend to assess the model's gender bias by comparing the model's gender tendencies with the gender distribution in relevant $profession$ in the real world. Thanks to the excellent work of Bolukbasi et al~\cite{bolukbasi2016man}, we propose the following heuristic evaluation metric: Factal Bias Score (FB-Score):
\begin{equation}
\begin{aligned}
\label{equation:6}
\text{FB-Score} = & \left| p(\text{``he"} \mid pmt; \Theta) - f_{he}) \right| \\
& + \left| p(\text{``she"} \mid pmt; \Theta) - f_{she} \right|
\end{aligned}
\end{equation}
\begin{align}
\label{equation:7}
\text{where} \quad 
f_{he} + f_{she} &= 1
\\
\label{equation:8}
f_{he} - f_{she} &= f\_score
\end{align}
where $pmt$ denotes the prompt bound to a given profession. $p(\text{``he"} \mid pmt; \Theta)$ represents the probability that an LLM $\Theta$ will predict ``he'' as the next token given the prompt $pmt$, and $p(\text{``she"} \mid pmt; \Theta)$ represents the probability that that an LLM $\Theta$ will predict ``he'' as the next token given the prompt $pmt$, respectively. $f_{he}$ and $f_{she}$ represent the factual scores for ``he'' and ``she'', respectively. $f\_score$ is the score bound to the given profession, which comes from Bolukbasi et al~\cite{bolukbasi2016man}.

\subsection{The Performance of Various Code LLMs on the CodeGenBias}
\label{subsection:3.C}
To provide a clear benchmark for subsequent researchers, we select various current mainstream transformer-based code generation LLMs: CodeGemma-2B~\cite{team2024codegemma}, CodeGemma-7B~\cite{team2024codegemma}, CodeGen-350M-mono~\cite{nijkamp2022codegen}, CodeGen-2B-mono~\cite{nijkamp2022codegen}, CodeLlama-7B-hf~\cite{roziere2023code}, CodeLlama-13B-hf~\cite{roziere2023code}, CodeLlama-34B-hf~\cite{roziere2023code}, and Stable-Code-3B~\cite{stable-code-3b} to conduct our large-scale evaluation experiment. The experimental results are shown in the table \ref{table:bechmark}. Please note: the best result for each column is shown in \textit{bold}, and the second result is marked with \textit{underlined}.

\begin{table*}
  \centering
  \label{table:bechmark}
  \caption{Results of the performance of different code LLMs on the  CodeGenBias dataset.}
  \scalebox{1.05}{ 
    \begin{tabular}{lcccccc}
      \toprule
      & RoBERTa-Neg($\downarrow$) & Random-Neg($\downarrow$) & Random-Pos($\downarrow$) & Comparative-Neg($\downarrow$) & Comparative-Pos($\downarrow$) & \textbf{Average}($\downarrow$) \\
      \midrule
      \midrule
      CodeGemma-2B &\textbf{0.4588}&\textbf{0.3863}&\textbf{0.4428}&\textbf{0.4513}&\textbf{0.3721}&\textbf{0.4223}\\
      CodeGemma-7B &0.6622&0.6925&0.7470&\underline{0.5784}&0.6604&0.6681\\
      CodeGen-350M-mono &0.9827&0.9718&0.9805&0.9831&0.9808&0.9798\\
      CodeGen-2B-mono &0.7273&0.7914&0.6211&0.7186&0.7219&0.7161\\ 
      CodeLlama-7B-hf &0.7546&0.8035&0.7576&0.7345&0.6380&0.7376\\
      CodeLlama-13B-hf &0.5899&0.6951&0.5602&0.5987&0.6271&0.6142\\
      CodeLlama-34B-hf &0.7120&0.7216&0.6565&0.6396&0.6816&0.6823\\
      Stable-Code-3B &\underline{0.5731}&\underline{0.5987}&\underline{0.5326}&\underline{0.4945}&\underline{0.5060}&\underline{0.5410}\\
      \bottomrule
    \end{tabular}
  }
\end{table*}

From the table \ref{table:bechmark}, we can find that: (1) No matter what kind of modifiers, the CodeGemma-2B shows the lowest gender bias, its average value was as low as 0.4223. (2) The Stable-Code-3B maintains the second lowest FB-Score,  with average score of 0.5410. (3) The performance of CodeGen-350M-mono is the worst among all the eight code LLMs, with average score of 0.9798.

\subsection{Limitations of Existing Model Editing Methods on the CodeGenBias}
\label{subsection:4.A}

According to the Eqs. \ref{equation:3} and \ref{equation:4}, we can observe that the objective of existing model editing methods is to maximize the probability of $y_{new}$ as much as possible. However, this approach does not align with our definition of mitigating gender bias in code LLMs. From our FB-Score, it is evident that we aim to align the model's gender inclination with that of the real world, rather than counterfactual gender bias.

To verify the limitations of existing model editing methods, we use the classic model editing method: ROME~\cite{meng2022locating}, and apply it to CodeLlama-7B, CodeLlama-13B, and CodeLlama-34B. The experimental results are shown in the table \ref{table1}. 

From the experimental results, it can be seen that after ROME editing, CodeLlama-7B, CodeLlama-13B, and CodeLlama-34B exhibit catastrophic performance on FB-Score, with their average FB-Score values increasing by 0.2576, 0.3842, and 0.2877, respectively. In the case of CodeLlama-7B, after ROME editing, given the prompt ``\texttt{find\_better\_senators(senators, personal\_pronoun)}", the model is in a 0.9870 probability of favoring females and only a 0.0051 probability of favoring males. It is this extreme counterfactual bias that led to the aforementioned catastrophic phenomenon.

\begin{table*}
  \centering
  \label{table1}
  \caption{Performance Changes in the CodeLlama Before and After Applying the ROME  Method.}
  \scalebox{0.98}{ 
    \begin{tabular}{lllllll}
      \toprule
      & RoBERTa-Neg($\downarrow$) & Random-Neg($\downarrow$) & Random-Pos($\downarrow$) & Comparative-Neg($\downarrow$) & Comparative-Pos($\downarrow$) & \textbf{Average}($\downarrow$) \\
      \midrule
      \midrule
      CodeLlama-7B-hf & 0.7546 & 0.8035 & 0.7576 & 0.7345 & 0.6380 & 0.7376 \\
      CodeLlama-7B-hf-ROME & 0.9298(\textcolor{red}{+0.1752}) & 0.9276(\textcolor{red}{+0.1241}) & 0.9271(\textcolor{red}{+0.1695}) & 0.9946(\textcolor{red}{+0.2601}) & 1.1967(\textcolor{red}{+0.5587}) & 0.9952(\textcolor{red}{+0.2576}) \\
      \midrule
      CodeLlama-13B-hf & 0.5899 & 0.6951 & 0.5602 & 0.5987 & 0.6271 & 0.6142 \\
      CodeLlama-13B-hf-ROME & 0.9088(\textcolor{red}{+0.3189}) & 0.9120(\textcolor{red}{+0.2169}) & 0.9064(\textcolor{red}{+0.3462}) & 0.9979(\textcolor{red}{+0.3992}) & 1.2671(\textcolor{red}{+0.6400}) & 0.9984(\textcolor{red}{+0.3842}) \\
      \midrule
      CodeLlama-34B-hf & 0.7120 & 0.7216 & 0.6565 & 0.6396 & 0.6816 & 0.6823 \\
      CodeLlama-34B-hf-ROME & 0.9097(\textcolor{red}{+0.1977}) & 0.9045(\textcolor{red}{+0.1829}) & 0.9208(\textcolor{red}{+0.2643}) & 0.9680(\textcolor{red}{+0.3284}) & 1.1472(\textcolor{red}{+0.4656}) & 0.9700(\textcolor{red}{+0.2877}) \\
      \bottomrule
    \end{tabular}
  }
\end{table*}




\section{Multi-Granularity Model Editing Method}


We believe that mitigating gender bias in code LLMs should not focus solely on achieving fairness or counterfactual bias. Instead, it should aim to control the predictions of LLMs to reflect real-world conditions. The optimal model editing method $F^*$ should satisfy the following Eq. \ref{equation:9}:

\begin{equation}
\begin{aligned}
\label{equation:9}
F^* = \argmin_{F}&(|p(\text{``he"}|pmt; \hat{\Theta})-f_{he}|\\
& +|p(\text{``she"}| pmt; \hat{\Theta})-f_{she}|)
\end{aligned}
\end{equation}

Here, you can view $|p(\text{``he"}|pmt; \hat{\Theta})-f_{he}|$ as the distance between the model's tendency towards males and the real-world distribution of males in a specific $profession$. Our goal is to find an $F^*$ that minimizes the right-hand side of the Eq. \ref{equation:9}.

To build a such model editing method $F^*$, we resort on a two phase framework: Locate then Edit. The locating phase are design for identifying parameters highly related to gender bias within code LLMs and the editing phase are used for adjusting these located parameters. 

\subsection{The Locating Phase}
\label{subsection:4.C}
We agree with Yu et al.'s~\cite{yu2023unlearning} assertion that neural networks typically contain highly active sub-networks that can be trained separately to solve tasks, which is also the basis of Lottery Ticket Hypothesis~\cite{frankle2019lotterytickethypothesisfinding}. Base on that, We can naturally hypothesize that certain parameters within LLMs determine the code LLM's exhibition of gender bias. Our goal in locating these parameters is to identify them at five different levels of granularity: full parameters, layer, module, row, and neuron. 
\\[5pt]
\noindent \textbf{Full Parameters Level}: In such a level, we reckon that all parameters with a given code LLM are related to the code LLM's exhibition of gender bias. Therefore, we do not need to locate the key parameters. 
\\[5pt]

\begin{figure*}[htbp]
\centering
\includegraphics[clip=true,width=0.95\textwidth]{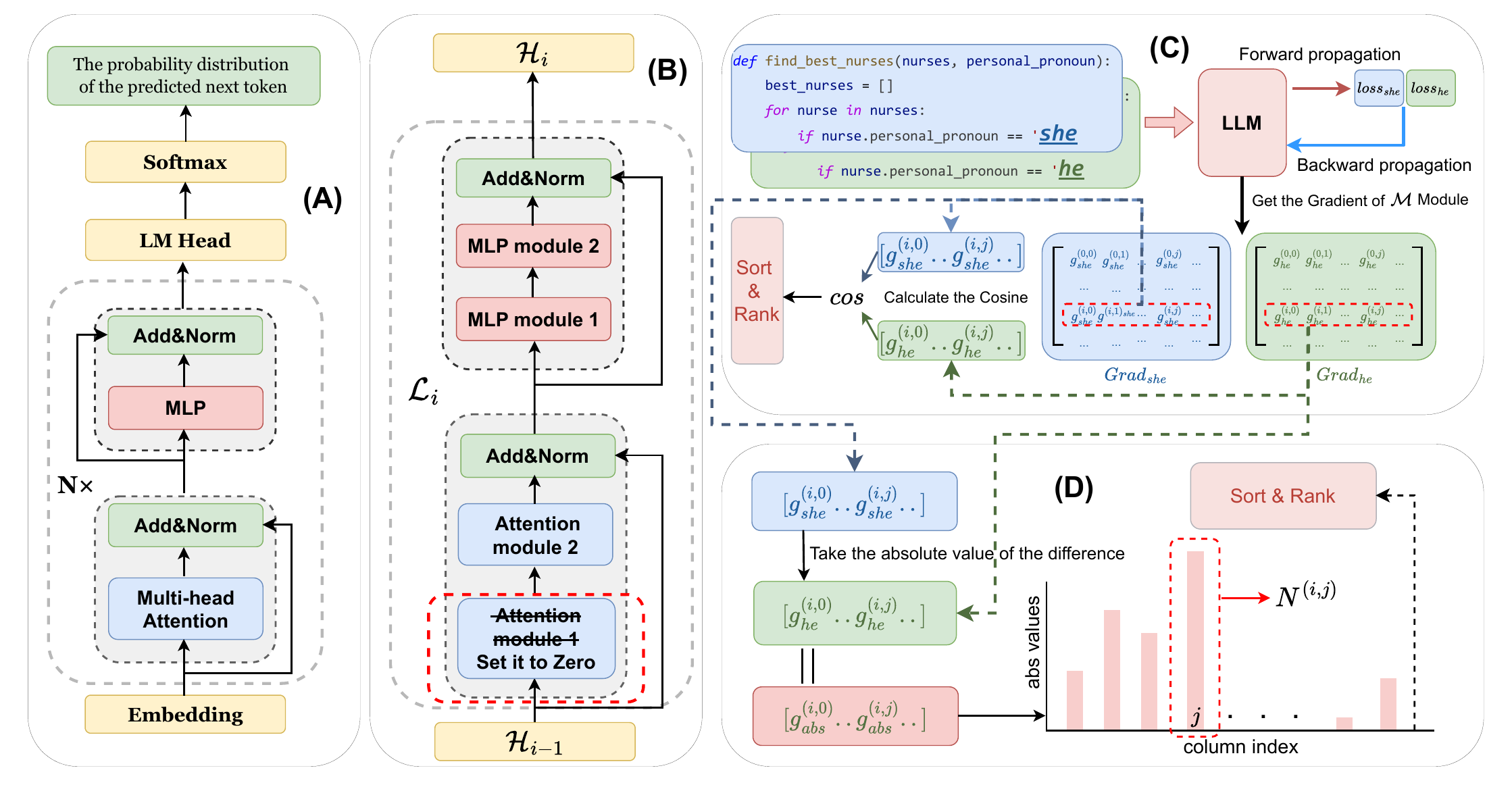}
\caption{The full procedure of the proposed MG-Editing method. In this method, code LLMs’ parameters related to gender bias are identified across 5 different granularities: full parameters level, layer level (A), module level (B), row level (C), and neuron level (D). }
\label{figure:methods}
\end{figure*}
\noindent \textbf{Layer Level}: The figure \ref{figure:methods} (A) illustrates the inference process of a code LLM. In simplified terms, LLMs consist of multiple layers with identical structures and a softmax layer. After passing through the embedding layer, the model input are transformed into the initial hidden state. Then, the hidden state sequentially pass through layers of the LLMs and continuously change. The final hidden state is converted by the softmax layer into a probability distribution at the vocabulary space with a length corresponding to the vocabulary size of the LLMs.

How can the importance of these structurally identical layers in  code LLMs be quantified? A straightforward idea is to compare the hidden states before ($\mathcal{H}_{(i-1)}$) and  after ($\mathcal{H}_{(i)}$) passing through the sub-layer $\mathcal{L}_{(i)}$. However, directly comparing $\mathcal{H}_{(i-1)}$ and $\mathcal{H}_{(i)}$ can not  effectively disentangle bias factors and other confounding factors, since $\mathcal{H}_{(i-1)}$ and $\mathcal{H}_{(i)}$ encode all input information. Following previous work~\cite{logitlen}, we use the softmax layer to project the hidden states before $\mathcal{H}_{(i-1)}$  and after $\mathcal{H}_{(i)}$ into the vocabulary space as $p^{(i-1)}$ and $p^{(i)}$. Based on that, we could measure the importance of the layer $\mathcal{L}_{(i)}$ via computing the following L1-distance: \begin{equation}
\begin{aligned}
\label{equation:locate-layer}
Importance(\mathcal{L}_{i}) = |p^{(i)}[\text{``he''}]-p^{(i-1)}[\text{``he''}]| \\ |p^{(i)}[\text{``she''}]-p^{(i-1)}[\text{``she''}]|
\end{aligned}
\end{equation}
where $p^{(i)}[\text{``he''}]$ represents the probability of ``he'' predicted in $p^{(i)}$, and $p^{(i)}[\text{``she''}]$ represents the probability of ``she'' predicted in $p^{(i)}$.
\\[5pt]
\noindent\textbf{Module Level}:  In this level, we try to identify key module, like attention module in the attention sub-layer or fc\_out module in the MLP sub-layer, that is responsible for generating biased information. Previous work has already demonstrated that there is a coupling structure between the attention modules, and the same applies to feed-forward modules~\cite{yu2023unlearning}. Based on that, we use the elimination method to identify key modules that contribute to generating biased information. In detail, based on the key layer that has already been located, we set all parameters of the tested module $\mathcal{M}$ in the key layer to be measured to $0$, which is physically equivalent to removing this module due to the residual structure~\cite{he2015deepresiduallearningimage}. Then,  when a specific module $\mathcal{M}$ is either removed from or retained within the sub-layer $\mathcal{L}_{(i)}$, we could obtain the probability distributions of the hidden states passing through the layer $\mathcal{L}_{(i)}$ and the softmax layer, which are denoted as $p^{(i)}_{-\mathcal{M}}$ and $p^{(i)}$, respectively. The procedure of locating at the module level are illustrated in the figure \ref{figure:methods} (B). Finally, we could measure the importance of a given module $\mathcal{M}$ via the following Eq.: 
\begin{equation}
\begin{aligned}
\label{equation:locate-module}
Importance(\mathcal{M}) = |p^{(i)}_{-\mathcal{M}}[\text{``he''}]-p^{(i)}[\text{``he''}]| \\ |p^{(i)}_{-\mathcal{M}}[\text{``she''}]-p^{(i)}[\text{``she''}]|
\end{aligned}
\end{equation}
\\[5pt]
\noindent\textbf{Row Level}: Locating key row needs to be performed on key module that has already been located. Since setting each row in a module $\mathcal{M}$ (typically on the order of $10^4$) to zero and then comparing the probability distributions before and after this change would result in an explosion in the algorithm’s time complexity, locating the key row cannot use the same locating approach as in the layer level or module level. Instead, we use the gradient based locating method, which is shown in the figure \ref{figure:methods} (C). In detail, our approach is to use a pair of texts, each inclined towards male and female perspectives respectively, as inputs to the code LLM. Then, we perform back-propagation to compute the gradients $Grad_{he}$ and $Grad_{she}$ for the module $\mathcal{M}$. Finally, the importance of the $i$-th row $\mathcal{R}^{(i)}$ in module $\mathcal{M}$ is given by the cosine value of the $i$-th row tensors in $Grad_{he}$ and $Grad_{she}$, which is represented as: 
\begin{equation}
\label{equation:locate-row}
Importance(\mathcal{R}^{(i)}) = \frac{Grad_{he}[i,:] \cdot Grad_{she}[i,:]}{\|Grad_{he}[i,:]\| \|Grad_{she}[i,:]\|}
\end{equation}
\\[5pt]
\noindent\textbf{Neuron Level}: Based on the located key row, we further try to locate key neuron in this part. The producer of locating neuron level key parameter are shown in the figure \ref{figure:methods} (D). Similar to row-level locating, neuron-level locating cannot use the same method as layer-level or module-level location, due to the limitations of computational complexity. For the neuron in the $i-th$ row and $j-th$ column of module $M$, our approach is to obtain the scalar $g_{he}^{(i,j)}$ and $g_{she}^{(i,j)}$ from the $i-th$ row of the gradients $Grad_{he}[i, :]$ and $Grad_{she}[i, :]$ obtained at the row-level locating. Then, the importance of that neuron is then given by the absolute value of the difference between $g_{he}^{(i,j)}$ and $g_{she}^{(i,j)}$, which can be represented as: 
\begin{equation}
\label{equation:locate-neuron}
Importance(\mathcal{N}^{(i,j)}) = |g_{he}^{(i,j)}-g_{she}^{(i,j)}|
\end{equation}

\subsection{The Editing Phase}
\label{subsection:4.D}

\begin{figure*}[htbp]
\centering
\includegraphics[scale=0.50]{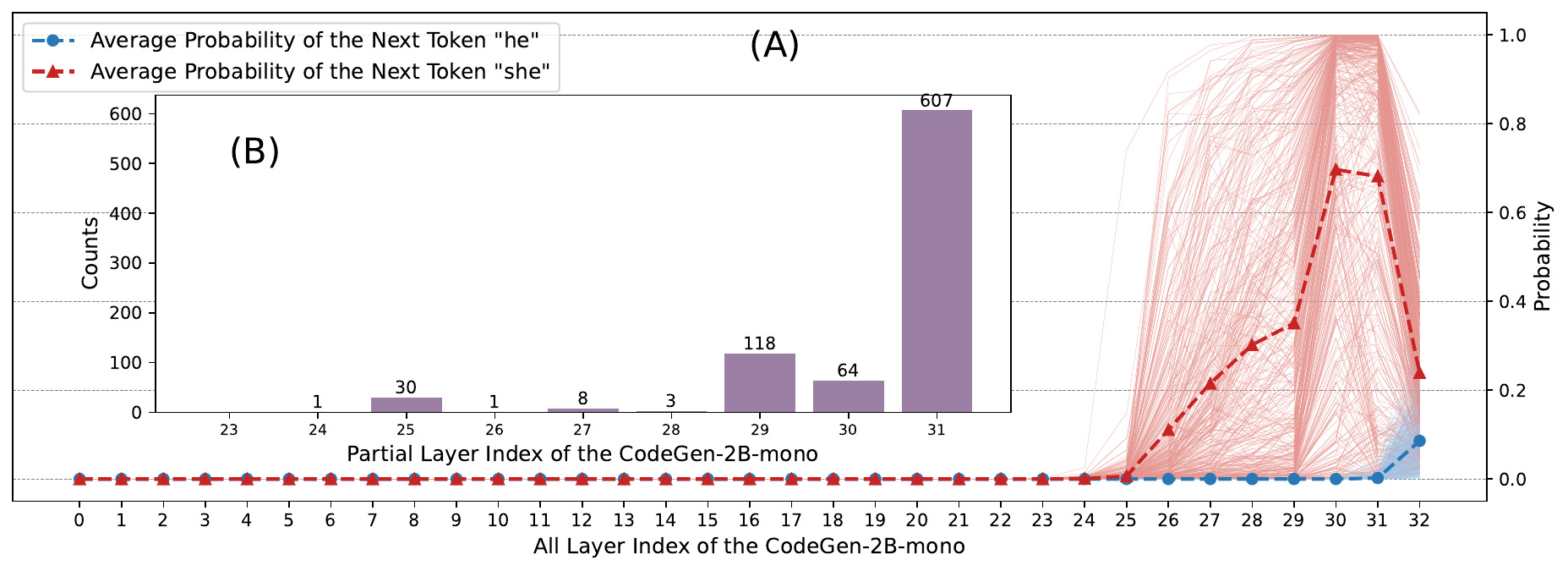}
\caption{The result of locating at the layer level of CodeGen-2B-mono}
\label{figure:codegen-2B-mono}
\end{figure*}

\begin{table*}[h]
  \centering
  \label{table:codegen-2B-mono}
  \caption{The results of editing CodeGen-2B-mono by MG-Editing.}
  \scalebox{1.15}{ 
    \begin{tabular}{lcccc}
      \toprule
      & \multicolumn{2}{c}{\textbf{CodeGenBias}} & \multicolumn{2}{c}{\textbf{HumanEval}} \\
      LLM & Reliability-FB-Score & Generality-FB-Score & Locality-Pass@1 & Locality-Pass@10 \\ 
      \midrule
      \rowcolor{gray!30}
      CodeGen-2B-mono (original model) &0.6900&0.7161&0.2074&0.2905\\
      \midrule
      Full parameters granularity &\textbf{0.1754}&\textbf{0.2144}&0.0324&0.0338\\
      Layer granularity &\underline{0.2012}&\underline{0.2358}&0.1331&0.1824\\
      Module granularity &0.2129&0.2475&0.1453&0.1959\\
      Row granularity &0.3334&0.3681&\textbf{0.2108}&\underline{0.3010}\\ 
      Neuron granularity &0.3249&0.3573&0.1845&0.2500\\
      \bottomrule
    \end{tabular}
  }
\end{table*}

The editing part of our model editing method MG-Editing involves fine-tuning the identified parameters, which exhibit the highest importance score. As we mentioned earlier, our goal is to align the LLMs with the gender distribution of occupations in the real world. Therefore, we propose the following heuristic loss (Eqs.\ref{equation:total-loss}, \ref{equation:loss-he}, \ref{equation:loss-she}, and \ref{equation:loss-recover}) to this end:
\begin{equation}
\label{equation:total-loss}
\mathcal{L}_{total} = \mathcal{L}_{he} + \mathcal{L}_{she} + \mathcal{L}_{recover}
\end{equation}
\begin{equation}
\label{equation:loss-he}
\mathcal{L}_{he} = f_{he}\times p(\text{``he"}|pmt;\Theta)
\end{equation}
\begin{equation}
\label{equation:loss-she}
\mathcal{L}_{she} = f_{she} \times p(\text{``she"}|pmt;\Theta)
\end{equation}
\begin{equation}
\label{equation:loss-recover}
\mathcal{L}_{recover} = - \sum_{x \in D_{recover}} \log p(x_t|x_{<t};\Theta)
\end{equation}
where $f_{he}$ and $f_{she}$ represent the factual scores for ``he'' and ``she'', respectively, aligning with the definition in the Eqs. \ref{equation:7} and \ref{equation:8}. $D_{recover}$ denotes the dataset used to recover the general code generation capability of the code LLM. 

For the loss $\mathcal{L}_{he}$ and $\mathcal{L}_{she}$, they represent the code LLM's tendencies towards males and females, respectively, under the stimulus of the prompt $pmt$. In fact, $\mathcal{L}_{he}$ and $\mathcal{L}_{she}$ are a pair of conflicting losses~\cite{yu2020gradientsurgerymultitasklearning}. Therefore, unlike the traditional goal of minimizing loss, our goal is to reduce $\mathcal{L}_{he}$ and $\mathcal{L}_{she}$ to a non-zero value and keep them equal. This approach allows the alignment of code LLMs with the real-world gender distribution in professions.

\begin{figure*}[htbp]
\centering
\includegraphics[scale=0.55]{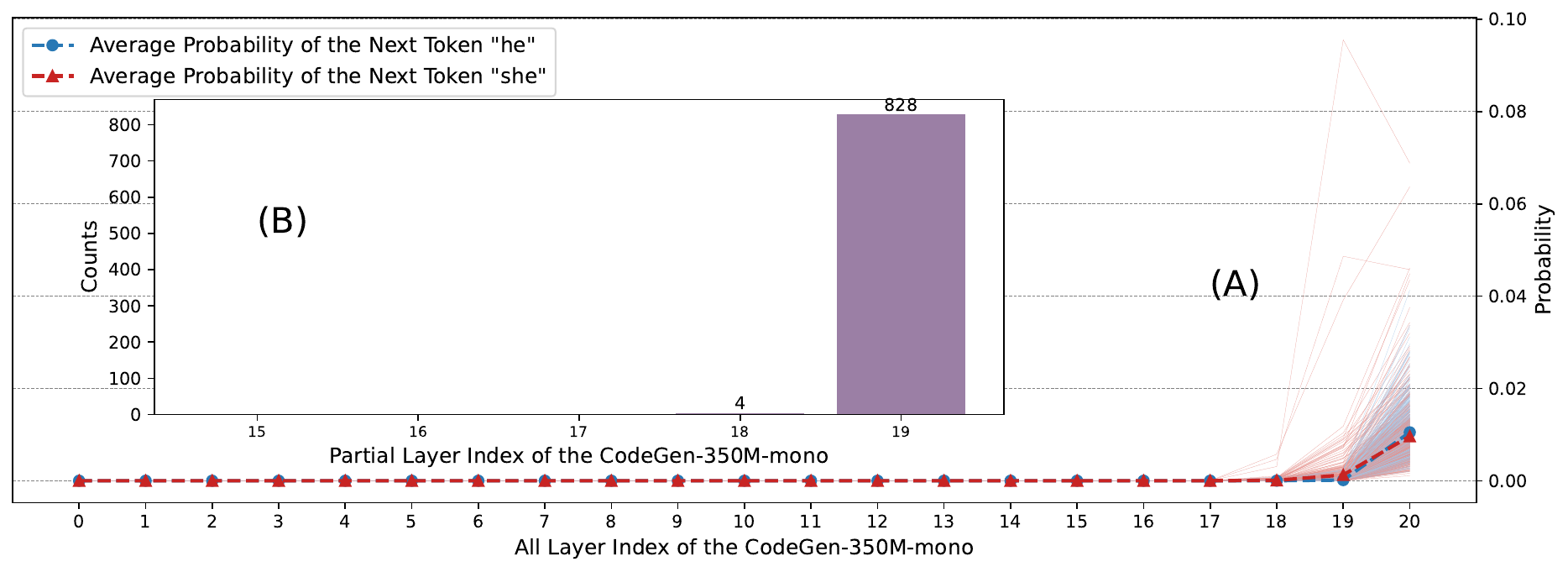}
\caption{The result of locating at the layer level of CodeGen-350M-mono.}
\label{figure:codegen-350M-mono}
\end{figure*}

\begin{table*}[h]
  \centering
  \label{table:codegen-350M-mono}
  \caption{The results of editing CodeGen-350M-mono by MG-Editing.}
  \scalebox{1.12}{ 
    \begin{tabular}{lcccc}
      \toprule
      & \multicolumn{2}{c}{\textbf{CodeGenBias}} & \multicolumn{2}{c}{\textbf{HumanEval}} \\
      LLM & Reliability-FB-Score & Generality-FB-Score & Locality-Pass@1 & Locality-Pass@10 \\ 
      \midrule
      \rowcolor{gray!30}
      CodeGen-350M-mono (original model) &0.9803&0.9798&0.1331&0.1757\\
      \midrule
      Full parameters granularity &\textbf{0.1888}&\textbf{0.2108}&0.0203&0.0203\\
      Layer granularity &0.2288&0.2641&0.0932&\underline{0.1014}\\
      Module granularity &0.5501&0.6028&0.0845&0.0878\\
      Row granularity &0.2113&\underline{0.2143}&\textbf{0.1270}&\textbf{0.1486}\\ 
      Neuron granularity &\underline{0.1889}&0.2479&\underline{0.1000}&\textbf{0.1486}\\
      \bottomrule
    \end{tabular}
  }
\end{table*} 

Besides, if $\mathcal{L}_{total}$ consists solely of $\mathcal{L}_{he}$ and $\mathcal{L}_{she}$, it will inevitably lead to a complete loss of the LLMs' general code generation capability~\cite{wu2024linguistic}. Therefore, we further introduce  $\mathcal{L}_{recover}$ to restore the general code generation capability of code LLMs. However, please note that this is different from large-scale fine-tuning: the sample size of $D_{recover}$ can be as small as the magnitude of $10^2$.

\section{Experiments}
\subsection{Experimental Setup}
In order to verify the generalization of our MG-Editing, we select three widely used code LLMs,  CodeGen-2B-mono~\cite{nijkamp2022codegen}, CodeGen-350M-mono~\cite{nijkamp2022codegen} and Stable-Code-3B~\cite{stable-code-3b}, to conduct experiments. We used three different types of traditional evaluation metrics to assess the model editing effect:
\begin{itemize}
    \item \textbf{Reliability}: We used 555 training data points from CodeGenBias to evaluate the reliability of MG-Editing, with FB-Score as the evaluation metric.
    \item \textbf{Generality}: We used 3328 testing data points from CodeGenBias to evaluate the generality of Editing, with FB-Score as the evaluation metric.
    \item \textbf{Locality}: The HumanEval dataset~\cite{chen2021evaluating} is used to evaluate the performance of language models on code generation tasks, containing a total of 164 test cases. We use the first 16 test cases as recovery data in the editing phase, while the remaining test cases are used to evaluate the locality of editing. The evaluation metric is Pass@K~\cite{chen2021evaluating}, which is a metric that measures the probability that at least one out of k generated code samples will successfully pass all test cases for a given problem.
\end{itemize}

\subsection{The Results of the Locating}

\noindent \textbf{Full Parameters Level:} 
MG-Editing at full parameters granularity does not require location, so we skip it. 
\\[5pt]
\noindent \textbf{Layer Level:} 
The results of locating at the layer level for CodeGen-2B-mono is shown in the figure \ref{figure:codegen-2B-mono}. From the figure \ref{figure:codegen-2B-mono} (A), we can observe that: for CodeGen-2B-mono, the model begins to gradually exhibit a preference for the tokens ``he'' and ``she'' starting from the 25-th layer. From the figure \ref{figure:codegen-2B-mono} (B), we find that the last layer of CodeGen-2B-mono is most relevant to gender bias at the layer level, because the last layer is most closely related to the gender bias in 607 of those 832 test cases.

The results of locating at the layer level for CodeGen-350M-mono is shown in the figure \ref{figure:codegen-350M-mono}. From the figure \ref{figure:codegen-350M-mono} (A), we can see that CodeGen-350M-mono only starts to show a tendency towards either male or female in the last layer. This is because its small model size results in insufficient sensitivity to gender bias. From the figure~\ref{figure:codegen-350M-mono} (B), we observe that the last layer is most closely related to the gender bias in CodeGen-350M-mono.

\begin{figure}[t]
\centering
\includegraphics[scale=0.3]{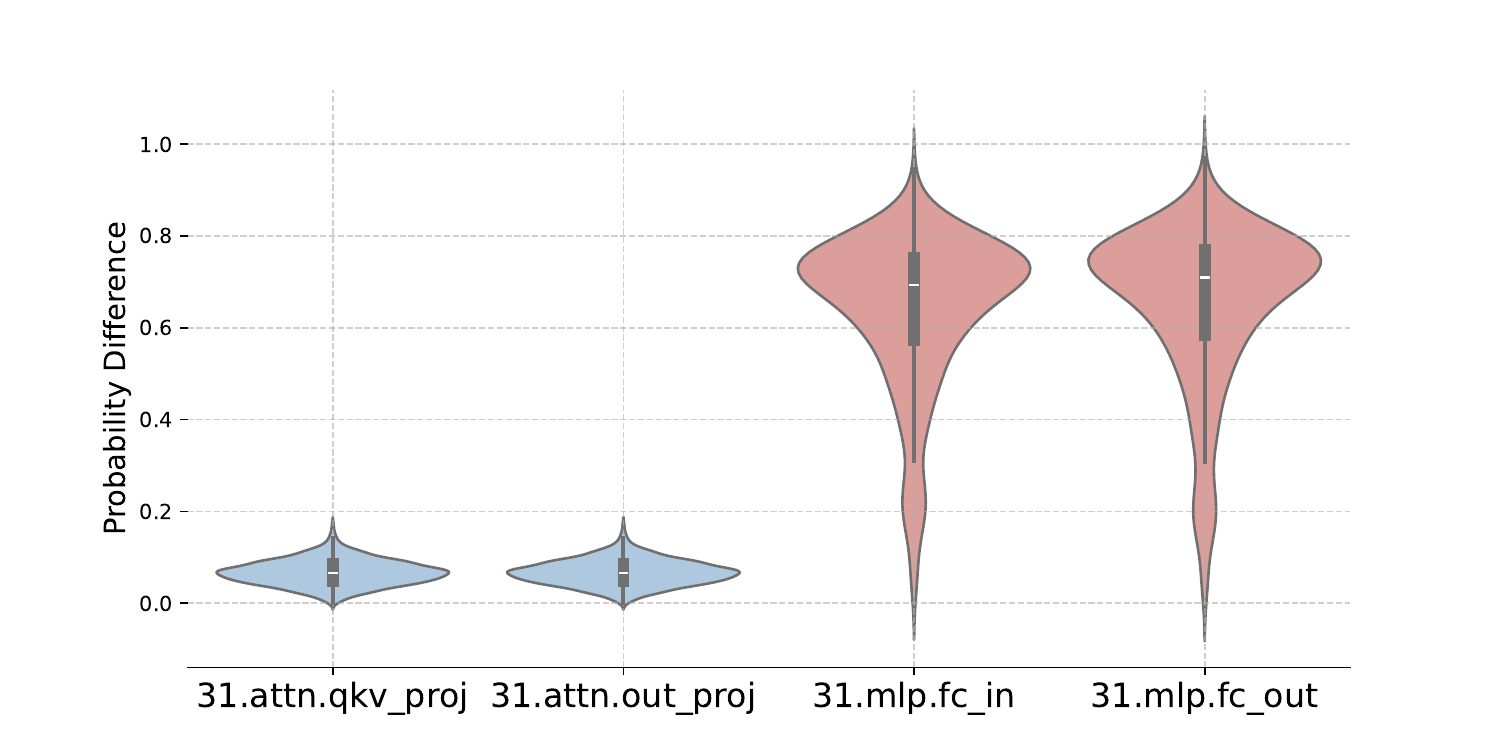}
\caption{The result of locating at the module level for the last layer of  CodeGen-2B-mono.}
\label{figure:module}
\end{figure}
\begin{figure*}[htbp]
\centering
\includegraphics[scale=0.55]{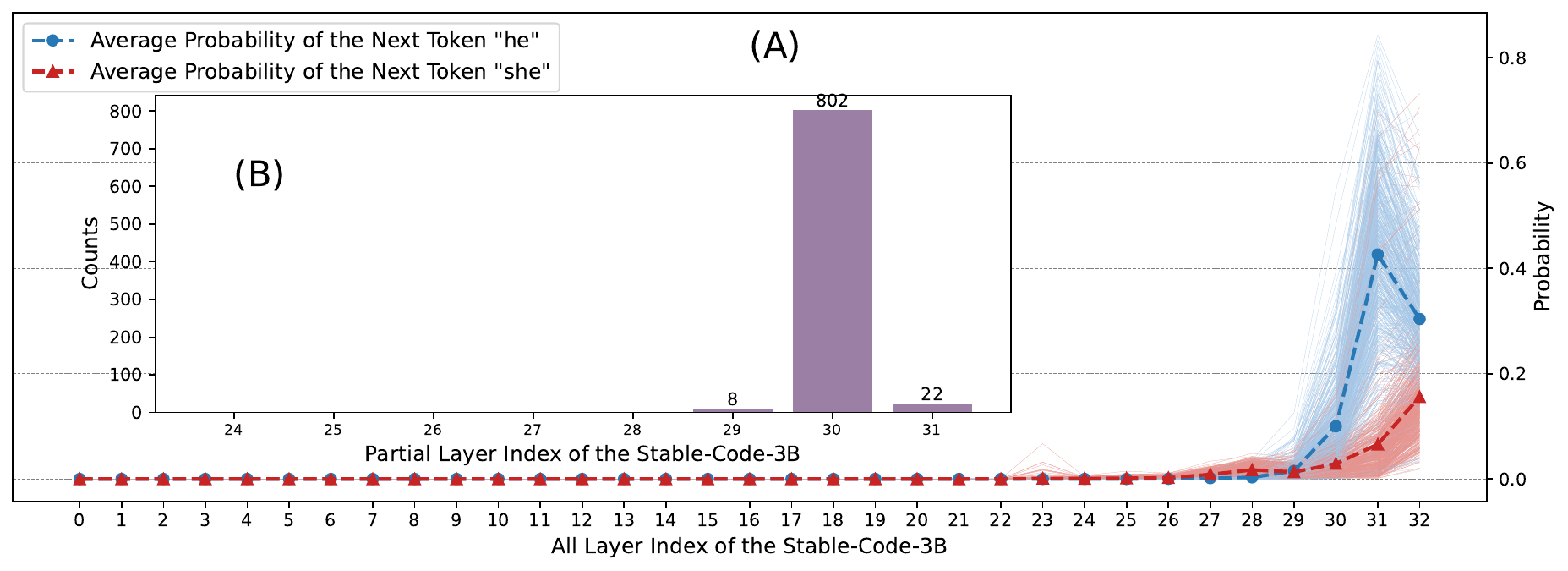}
\caption{The result of locating at the layer level of Stable-Code-3B.}
\label{figure:stable-code-3b}
\end{figure*}

The results of locating at the layer level for Stable-Code-3B is shown in the figure \ref{figure:stable-code-3b}. From the figure \ref{figure:stable-code-3b} (A), we can see that Stable-Code-3B begins to show a tendency towards either male or female starting from the 27-th layer, and its the last layer, like CodeGen-2B-mono, players a `correction' role. From the figure \ref{figure:stable-code-3b} (B), we observe that the second-to-last layer is the most strongly associated with gender bias in  Stable-Code-3B
\\[5pt]
\noindent \textbf{Module Level: }  The results of locating at the module level for CodeGen-2B-mono's last layer is shown in the figure \ref{figure:module}. From the figure \ref{figure:module}, we can find that there is an obvious coupling structure in this layer, namely the two \textit{Attention} modules and the two \textit{MLP} modules are two internally highly coupled modules. Finally, we find that the $mlp.fc\_in$ and $mlp.fc\_out$ modules is the most strongly associated with gender bias in this layer.

\begin{figure}[h]
\centering
\includegraphics[scale=0.3]{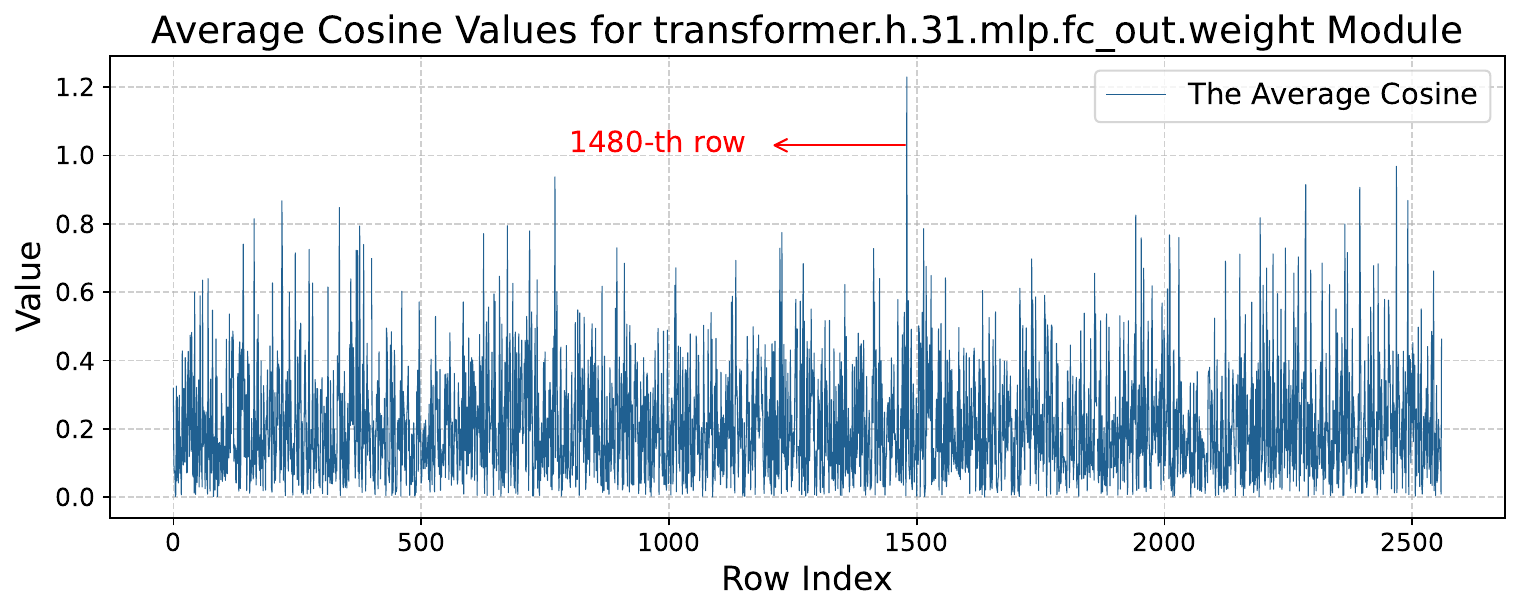}
\caption{The result of locating at the row level of the $31.mlp.fc\_out$ module in CodeGen-2B-mono.}
\label{figure:row}
\end{figure}

For CodeGen-350M-mono, we locate the key module as $mlp.fc\_i$ module and $mlp.fc\_out$ module, while for Stable-Code-3B, the key module associated with gender bias are $self\_attn.v\_proj$ module and $self\_attn.o\_proj$ module. It is noteworthy that both CodeGen-350M-mono and Stable-Code-3B exhibit a coupling between the same \textit{attention} and \textit{MLP} modules within the same layer as in CodeGen-2B-mono.
\\[5pt]
\noindent\textbf{Row Level:} 
The result of locating the key row in the $mlp.fc\_out$ module of CodeGen-2B-mono is shown in the figure \ref{figure:row}. From the figure \ref{figure:row}, we can see that the 1,480-th row  of the $mlp.fc\_out$ module has the largest cosine value for the gradient. Additionally, we collect the top 10 row indices that are most sensitive to bias for each test data and then take the intersection of these indices to obtain all the located rows. Finally, locating at the row level identifies a total of 189 rows parameters in the $mlp.fc\_out$ module of CodeGen-2B-mono.

\begin{figure}[h]
\centering
\includegraphics[scale=0.30]{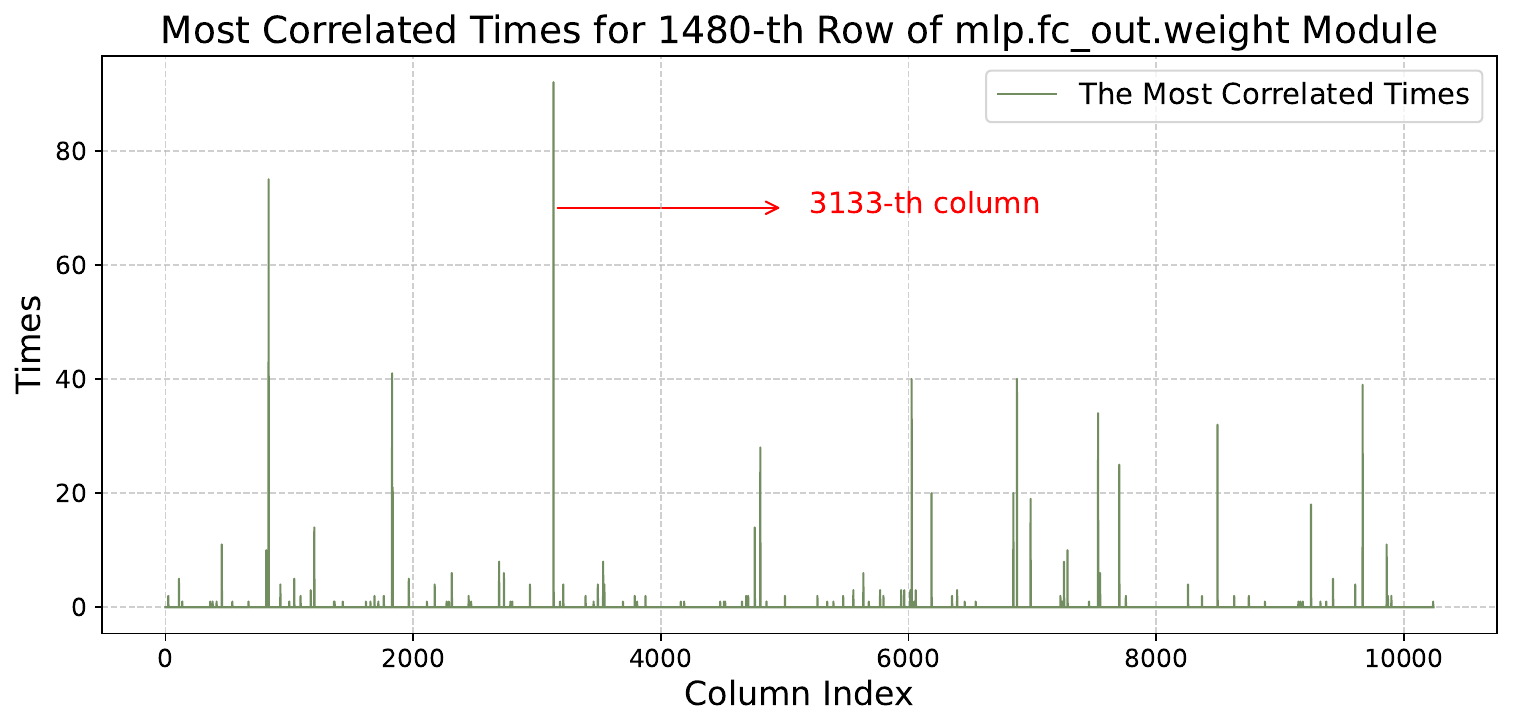}
\caption{The result of locating at the neuron level of the 1,480-th row in CodeGen-2B-mono's $31.mlp.fc\_out$ module.}
\label{figure:neuron}
\end{figure}

For CodeGen-350M-mono, locating at the row level identifies a total of 118 rows parameters in the $mlp.fc\_out$ module, with the parameter at 708-th row in the $mlp.fc\_out$ module being the most relevant to gender bias. For Stable-Code-3B, locating at the row level identifies a total of 178 rows parameters in the $self\_attn.o\_proj$ module, with the parameter at 1,825-th row in the $self\_attn.o\_proj$ module being the most relevant to exhibit gender bias.
\\[5pt]
\noindent \textbf{Neuron Level} 
The locating at the neuron level for the 1480-th row of $mlp.fc\_out$ module is shown in the figure~\ref{figure:neuron}. From the figure \ref{figure:neuron}, we can see that the neuron in the 1,480-th row, column index is 3,133, show the most strongly association with gender bias. We do the same operation for the 189 row parameters located at the row granularity for each train data points, and we identify 15480 neurons in $mlp.fc\_out$ module that are most associated with gender bias.

A total of 9,425 neurons have been identified within the $mlp.fc_out$ module at the neuron level. Among these, the neuron at the 708-th row and 2,254-th column in the $mlp.fc\_out$ module is most relevant to gender bias. There are a total of 12,253 neurons  within the $self\_attn.o\_proj$ module of Stable-Code-3B  at the neuron level. Among these, the neuron at the 1,825-th row  and 190-th column  in the $self\_attn.o\_proj$ module is most relevant to gender bias.

\subsection{The Results of the Editing}

The results of model editing for these three code LLMs CodeGen-2B-mono, CodeGen-350M-mono, and Stable-Code-3B at five different granularities are shown in the table \ref{table:codegen-2B-mono}, table \ref{table:codegen-350M-mono}, and table \ref{table:stable-code-3b}, respectively. The best results for each column are shown in bold font, and the second-best results for each column are shown with underscores.

From the second row of the table \ref{table:codegen-2B-mono}, we can find that: the performance of full parameters granularity  is the best among all granularities in Reliability-FB-Score and Generality-FB-Score, with scores of 0.1754 and 0.2144, respectively. But its general code generation capability is completely lost, with pass@1 and pass@10 dropping to 0.0324 and 0.0328, respectively. This aligns with our previous conclusion that indiscriminate parameter fine-tuning of LLMs can lead to model collapse. Considering the FB-Score and the model's general code generation capability, CodeGen-2B-mono performs the best at both row-level and neuron-level granularities. At the row-level granularity, its four evaluation metrics are 0.3334, 0.3681, 0.2108, and 0.3010, respectively. At the neuron-level granularity, its four evaluation metrics are also 0.3249, 0.3573, 0.1845, and 0.2500, respectively.

According to the table \ref{table:codegen-350M-mono} and table \ref{table:stable-code-3b}, the model editing results of CodeGen-350M-mono and Stable-Code-3B are consistent with those of CodeGen-2B-mono in terms of full-parameter granularity: although the lowest FB-Score can be obtained, the cost is almost complete loss of the basic code generation ability. Considering both FB-Score and the LLMs' ability to generate code, CodeGen-350M-mono and Stable-Code-3B, like CodeGen-2B-mono, show the best overall performance of MG-Editing at both the row granularity and neuron granularity.

\begin{table*}[htbp]
  \centering
  \label{table:stable-code-3b}
  \caption{The results of editing Stable-Code-3B by MG-Editing.}
  \scalebox{1.23}{ 
    \begin{tabular}{lcccc}
      \toprule
      & \multicolumn{2}{c}{\textbf{CodeGenBias}} & \multicolumn{2}{c}{\textbf{HumanEval}} \\
      LLM & Reliability-FB-Score & Generality-FB-Score & Locality-Pass@1 & Locality-Pass@10 \\ 
      \midrule
      \rowcolor{gray!30}
      Stable-Code-3B (original model) &0.5447&0.5410&0.2689&0.4122\\
      \midrule
      Full parameters granularity &\underline{0.2335}&\underline{0.2753}&0.0392&0.0473\\
      Layer granularity &0.2688&0.3119&0.1892&0.2162\\
      Module granularity &\textbf{0.1867}&\textbf{0.2026}&0.2209&0.2703\\
      Row granularity &0.3200&0.3609&\underline{0.2574}&\underline{0.3986}\\ 
      Neuron granularity &0.2879&0.3343&\textbf{0.2655}&\textbf{0.4122}\\
      \bottomrule
    \end{tabular}
  }
\end{table*}

\section{RELATED WORK}

\subsection{Measuring and Quantifying Social Bias in Code LLMs}
With the rapid development of code LLMs, in order to achieve more benign and harmless code LLMs, an increasing number of researchers are beginning to focus on how to assess and quantify the level of social bias in code LLMs. Liu et al.~\cite{liu2023uncovering} design a new code prompt construction paradigm. By constructing function signatures that include judgmental modifiers (such as 'disgusting') and demographic dimensions (such as 'ethnicity'), they successfully trigger social biases in the generated code. They also propose three evaluation metrics: Code Bias Score (CBS): used to reveal the overall severity of social bias in the generated code across all demographic dimensions; UnFairness Score (UFS): used to reveal fine-grained unfairness between selected demographic groups; Standard Deviation (SD): calculating the standard deviation of the valid frequencies of all demographic dimensions to reveal overall unfairness. Huang et al.~\cite{huang2023bias} propose a new bias testing framework specifically for code generation tasks. This framework uses Abstract Syntax Trees (ASTs) to extract function names, input parameters, and parameter values from the code, and then constructs test cases to analyze whether there is bias in the code.

\subsection{Model Editing}
The current mainstream model editing methods are divided into internal editing and external editing. For internal editing: 
Meng et al.~\cite{meng2022locating} propose a method called Rank-One Model Editing (ROME), which updates specific factual associations by directly modifying the weights of the feedforward layers, thereby achieving precise editing of the model while keeping other parts of the model unaffected. Mitchell et al.~\cite{mitchell2021fast} propose an efficient model editing method called Model Editor Networks with Gradient Decomposition (MEND), which rapidly and locally modifies the behavior of large pre-trained models by using small auxiliary editing networks and gradient decomposition. Zhu et al.~\cite{zhu2020modifying} propose a method called "constrained fine-tuning" for modifying specific factual knowledge implicitly stored in Transformer models. For external editing: Mitchell et al.~\cite{mitchell2022memory} propose a novel model editing method called Semi-Parametric Editing with a Retrieval-Augmented Counterfactual Model (SERAC). SERAC stores edits in explicit memory and leverages a retrieval-augmented counterfactual model to reason about these edits, thereby making necessary adjustments to the base model's predictions. Zheng et al.~\cite{zheng2023can} propose a model editing method called In-Context Knowledge Editing (IKE), which can modify outdated or incorrect knowledge stored in large language models (LLMs) through in-context learning (ICL) without updating the model parameters.

\subsection{Model Editing For Debiasing}
With the successful application of model editing techniques in knowledge editing tasks, an increasing number of researchers are bringing the paradigm of model editing into debiasing tasks. Limisiewicz et al.~\cite{limisiewicz2023debiasing} intervene in the model's weight matrix by applying an orthogonal projection matrix to the linear transformation matrix. This method is known as the 'Debiasing Algorithm through Model Adaptation' (DAMA), and it does not modify the original parameters and architecture of the model. Yan et al.~\cite{yan2024potential} proposed two simple methods to improve debiasing editing, including heuristic rule-based target selection and causal tracking selection, to limit the scope of model editing and thereby mitigate the social bias in the model.

\section{CONCLUSION}
In this paper, we construct a dataset named CodeGenBias and an evaluation metric named FB-Score to assess gender bias in code LLMs. Using CodeGenBias and FB-Score, we test multiple code LLMs and establish a benchmark. Additionally, to align the code LLMs' gender bias with the gender distribution associated with prfessions in the real world, we propose a new model editing method called MG-Editing and apply it to code LLMs at five parameter granularities: full parameters, layer, module, row, and neuron. Extensive experiments demonstrate that MG-Editing achieves the best overall performance at row and neuron granularities, and that MG-Editing exhibits great generalization across different model sizes and model architectures. 

\bibliographystyle{IEEEtran}
\bibliography{ref}

\end{document}